\documentclass[aps,twocolumn,pra,superscriptaddress,amsmath,showpacs,tightenlines]{revtex4-1}
\usepackage{amssymb}
\usepackage{amsmath}
\usepackage{graphicx}
\usepackage{subfigure}
\usepackage{natbib}
\usepackage{epsfig}
\usepackage{amsfonts}
\usepackage{mathrsfs}
\usepackage{CJK}
\usepackage{xcolor}
\usepackage[toc,page,title,titletoc,header]{appendix}

\begin{document}
\title{ Giant emitter magnetometer}
\author{Xiaojun Zhang}
\affiliation{Center for Quantum Sciences and School of Physics, Northeast Normal University,
Changchun 130024, China}
\author{Xiang Guo}
\affiliation{Center for Quantum Sciences and School of Physics, Northeast Normal University,
Changchun 130024, China}
\author{Zhihai Wang}
\email{wangzh761@nenu.edu.cn}
\affiliation{Center for Quantum Sciences and School of Physics, Northeast Normal University,
Changchun 130024, China}

\begin{abstract}
Leveraging the sensitive dependence of a giant atom's radiation rate on its frequency~[A. F. Kockum, $et~al$., Phys. Rev. A {\bf90}, 013837 (2014)], we propose an effective magnetometer model based on single giant emitter. In this model, the emitter's frequency is proportional to the applied bias magnetic field. The self-interference effect causes the slope of the dissipation spectrum to vary linearly with the number of emitter-coupling points. The giant emitter magnetometer achieves a sensitivity as high as $10^{-8}-10^{-9}\,{\rm T/\sqrt{Hz}}$, demonstrating the significant advantages of the self-interference effect compared to small emitters. We hope our proposal will expand the applications of giant emitters in precision measurement and magnetometry.
\end{abstract}
%\pacs{42.50.Pq, 03.67.Lx, 42.50.Dv}
\maketitle
\section{introduction}

The precise measurement of magnetic fields is essential in fundamental scientific research, including geological exploration~\cite{ge,ge1}, aerospace applications~\cite{tian}, and biomedical imaging~\cite{biomedical}.
Optical magnetometers~(OM)~\cite{OMr,OMr1,OMr2,OMr3} have been extensively studied due to their advantages of high reliability, compact size, and low cost.
The working principle of an OM involves placing atoms with a specific spin state into a magnetic field. The interaction with the magnetic field alters the spin state of the atoms, inducing Larmor precession. The precession frequency is directly proportional to the magnetic field strength. By applying probe light to detect changes in the spin state, the magnetic field strength can be accurately measured.

Historically, the working substance of optical magnetometers consisted of natural atoms. Natural atoms interact with light fields, and because their size is much smaller than the wavelength of the light field, they can be approximated as point under the well-known dipole approximation.
This approximation remained valid until 2014, when the coupling between the transmon and surface acoustic waves was experimentally demonstrated~\cite{S1}, breaking the traditional framework. Due to its size being comparable to the wavelength of surface acoustic waves, the transmon can achieve multi-point coupling. Such transmons are referred to as giant atoms.
The multi-point coupling of giant atoms with their environment induces self-interference effect, giving rise to many intriguing physical phenomena. These include decoherence-free interactions~\cite{AF2018,AC2020,DL2023PRA}, frequency-dependent relaxation~\cite{ar2022,Lambpra14}, non-exponential decay~\cite{NM1,NM2,kim2023,guonjp}, chiral radiation~\cite{XW2021,XW2022q,guo,AS2022,FR2024,DL2023,prx2023}, phase-controlled entanglement~\cite{Jieqiao,Jieqiao1,PRL2023,weng}, and retardation effects~\cite{chengre,LG2017}.

In addition to using superconducting qubits to achieve the giant atom configuration~\cite{sq1,sq2,sq3,sq4}, the giant emitter configuration, based on the coupling of a ferromagnetic spin ensemble (yttrium iron garnet, YIG) sphere with bent waveguide, has also been experimentally realized~\cite{mq}.
The magneton mode of YIG spheres corresponds to the collective spin excitation mode. Quantum control of a single magneton in a YIG sphere has been successfully demonstrated~\cite{smq1,smq2,smq3,smq4}. When considering only single excitations, the giant emitter formed by YIG spheres can be modeled as a two-level emitter. The realization of such a giant emitter opens up a new platform for OM.

By combining the excellent controllability of YIG spheres with the self-interference effect of giant emitters, we propose a magnetometer operating in the microwave domain. As shown in Fig.~\ref{schem}(a), the YIG sphere, which serves as the working substance, couples to a bent waveguide at multiple points. The microwave photons emitted in the waveguide are reflected back and forth between the coupling points, resulting in the interference effect. Consequently, the dissipation rate of the YIG sphere becomes sensitive to its resonant frequency, which is determined by the applied bias magnetic field (MF). Therefore, the system provides an ideal platform for MF sensing.

We determine the optimal measurement point within the framework of the master equation and obtain the sensitivity with exact numerical calculation based on the full Hamiltonian of YIG-waveguide coupled system. Our results show that a sensitivity of $10^{-8}-10^{-9}\, {\rm T/\sqrt{Hz}}$ can be achieved, representing an order of magnitude enhancement compared to the small emitter counterpart.

The rest of the paper is organized as follows. In Sec.~\ref{model1}, we illustrate our model
of giant YIG and the dynamics within and beyond the Markovian approximation. In Sec.~\ref{meter}, we discuss the sensitivity of the proposed magnetometer and the performance of small emitters are investigated in Sec.~\ref{small}. We finally reach the brief conclusion in Sec.~\ref{con}. The detailed derivation can be found in the appendixes.

\section{model}
\label{model1}
\begin{figure}[t]
  \centering
  \includegraphics[width=7cm]{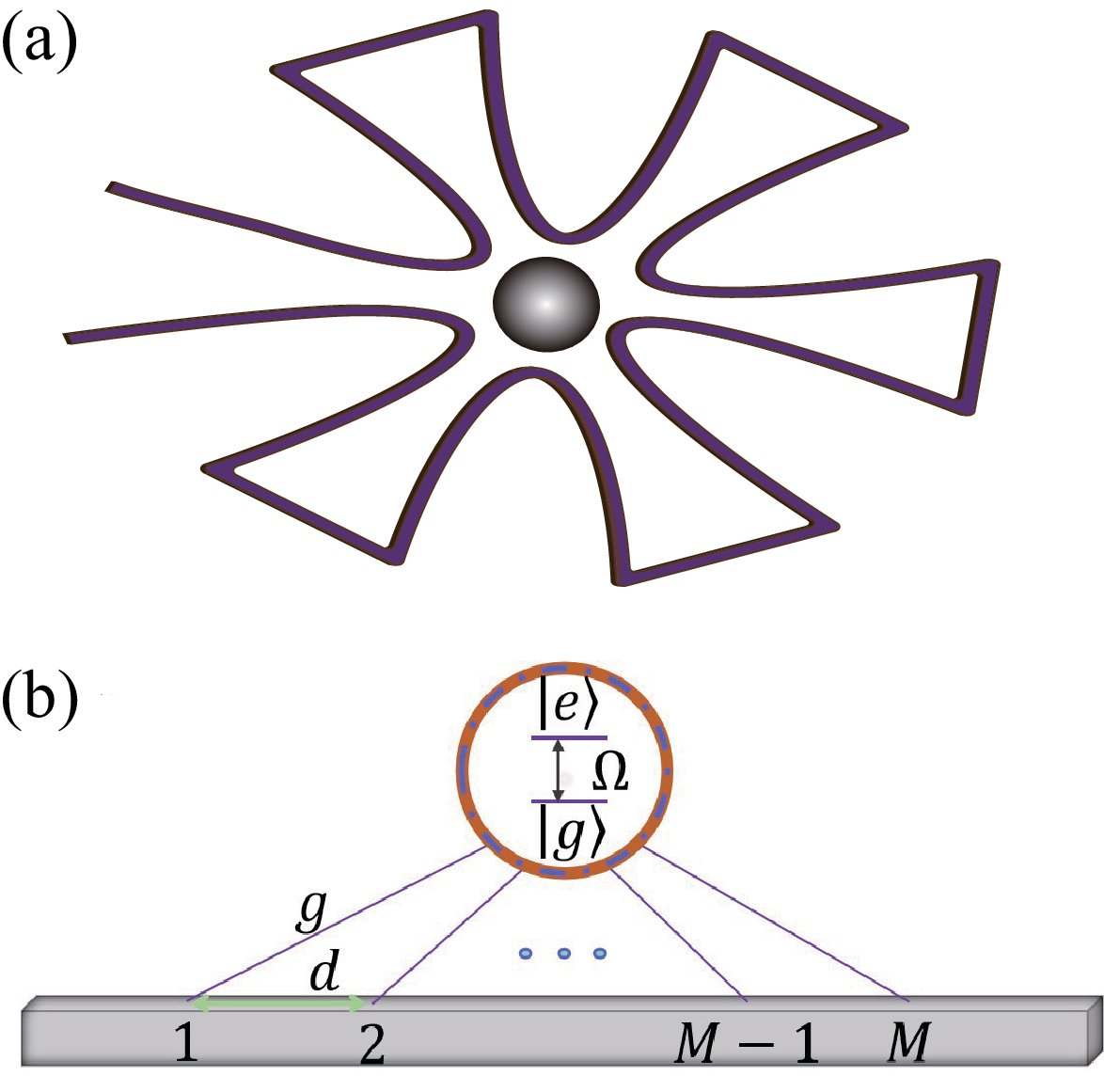}
  \caption{(a)~The implementation of the giant emitter, where black part is the YIG sphere and purple part is the bent waveguide. (b)~Sketch of the waveguide QED setup. A giant emitter
is coupled to the waveguide. There are $M$ coupling sites and the distance between the nearest coupling points is equal to $d$.}\label{schem}
\end{figure}
As illustrated in Fig.~\ref{schem}(b), the model we consider consists of a giant emitter coupled to a waveguide at $M$ coupling points. As discussed below, the emitter can be implemented using a magnon, while the waveguide is realized with microstrip circuits. This system provides an ideal platform for performing quantum metrology of external magnetic fields, which determine the resonant frequency of the magnon.

The Hamiltonian of the emitter-waveguide coupled system
is $H=H_A+H_W+H_I$ where
\begin{eqnarray}
H_A&=&\Omega|e\rangle\langle e|\nonumber\\
H_W&=&\int_{-\infty}^{+\infty} v |k|a_{k}^{\dagger}a_k dk\nonumber\\
H_I&=&\frac{g}{\sqrt{2\pi}}\sum_{n=1}^{M}\int_{-\infty}^{+\infty}dk(\sigma_+a_k e^{iknd}+{\rm H.c.}).\label{Ha}
\end{eqnarray}
 Here, $\Omega$ is the transition frequency of the giant emitter between its ground state $|g\rangle$ and excited state $|e\rangle$.
 $\sigma_+=|e\rangle\langle g|$ is the raising operator.  $k$ and $v$ is the wave vector and the group velocity of photons in the
 waveguide. $a_k$ is the photon annihilation operator of the $k$th mode in the waveguide. The coupling strength between the giant
 emitter and each coupling point is $g$. For the convenience of discussion, we define the total coupling strength between the emitter and the waveguide as $G=Mg$. We have assumed that the distance between arbitrary two nearest coupling points is $d$. In this paper, we set the length $d$ and the wave vector $k$ to be dimensionless,  so that the group velocity possesses the same dimension as the frequency.

 \subsection{Markovian approximation}

Under the Markovian approximation, where the waveguide is treated as the environment, the dynamics of the emitter is governed by the master equation (see Appendix~\ref{mas} for details)

\begin{eqnarray}
\frac{d\rho}{dt}&=&-i[(\Omega+L)|e\rangle\langle e|,\rho]+\frac{R}{2}D(\sigma_-)\rho\\
R&=&\frac{2g^2}{v}\sum_{n,l=1}^{M}\cos [|n-l|\phi]\label{cp1}\\
L&=&\frac{g^2}{v}\sum_{n,l=1}^M\sin [|n-l|\phi]\label{cp2}
\end{eqnarray}
where $D(O)\rho = 2 O\rho { O}^\dagger - \{{O}^\dagger O, \rho\}$, and $R$ and $L$ represent the effective decay rate and the Lamb shift induced by the waveguide environment, respectively. Here, $\phi = \Omega d/v$ denotes the phase accumulated by photons between adjacent coupling points of the giant emitter.
It is evident that the decay rate $R$ and Lamb shift $L$ are highly sensitive to the emitter's frequency $\Omega$, which is a unique feature of the giant emitter setup. This sensitivity enables us to use the emitter's population as an observable to perform metrology on its frequency and, by extension, the parameters that determine the frequency. For example, in the magnon system, the magnetic field can be measured, as discussed in the next section.

Before proceeding, we aim to evaluate the performance of our setup and identify the optimal parameter regime for metrology. Using the emitter's population in the excited state, $P_e = \langle |e\rangle\langle e| \rangle$, as the observable, the variance of the transition frequency, $\Delta^2 \Omega$, can be determined via the error transfer formula, which is expressed as~\cite{CG2024}

\begin{eqnarray}
\Delta^2\Omega=\frac{\langle(\sigma_+\sigma_-)^2\rangle-
\langle\sigma_+\sigma_-\rangle^2}{(\frac{\partial
\langle\sigma_+\sigma_-\rangle}{\partial \Omega})^2}
=\frac{Pe(1-Pe)}{(\frac{\partial Pe}{\partial \Omega})^2}.\label{error}
\end{eqnarray}
Governed by the master equation, emitter's population obeys the exponential decay of $P_e=\exp(-Rt)$, and
the classical Fisher information (CFI) associate with the population measurement is
\begin{eqnarray}
\Delta^2\Omega=\frac{e^{Rt}-1}{t^2(\frac{\partial R}{\partial \Omega})^2}.\label{F}
\end{eqnarray}

\begin{figure}
  \centering
  \includegraphics[width=8cm]{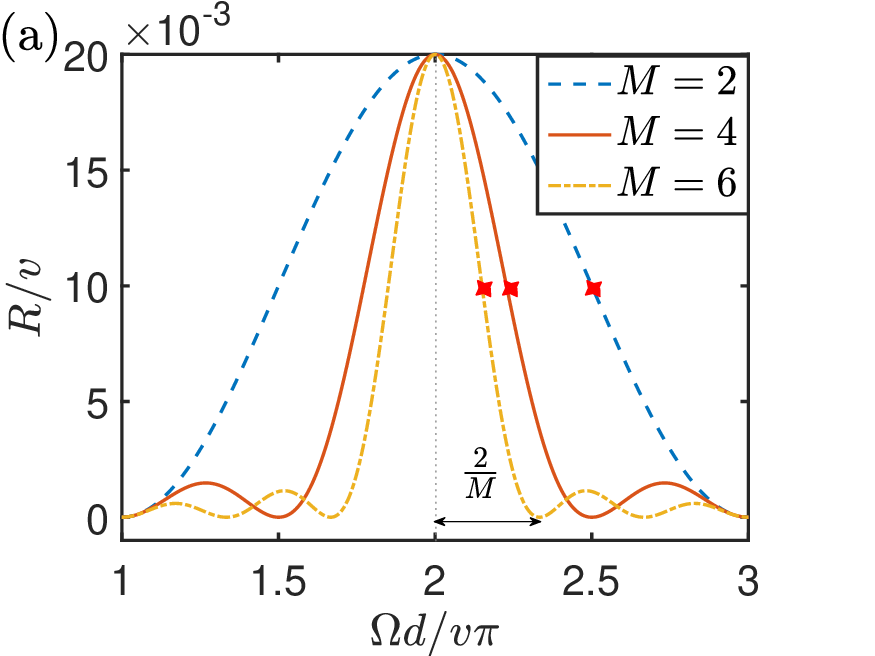}
  \includegraphics[width=8cm]{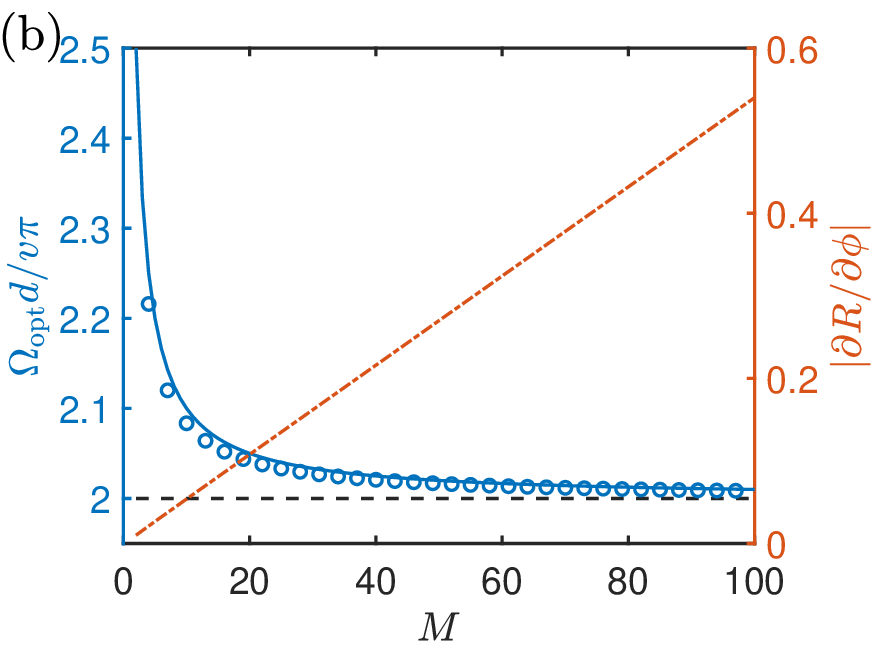}
  \caption{(a) Dissipative spectrum corresponding to different $M$. (b) The orange dot dash line represents the change of the maximum slope with $M$. The blue solid line represents the change of $\phi/\pi=\Omega_{\rm{opt}} d/(v\pi)$ with $M$ corresponding to the maximum slope of the coupling spectrum. The blue dotted line corresponds to $\phi/\pi$ at the peak of the coupling spectrum.  For (a,b), coupling strength $G=0.1V$. }\label{couple}
\end{figure}

Therefore, the slope $\partial R/\partial \Omega$ of the curve $R$ versus $\Omega$ plays a crucial role in suppressing fluctuations and enhancing the precision of parameter metrology. To identify the ideal working point for metrology, we plot the decay rate $R$ as a function of the emitter's frequency in Fig.~\ref{couple}(a) for different values of $M$.
The decay rate $R$ is a periodic function of the phase $\phi = \Omega d/v$ with a period of $2\pi$, as explicitly shown in Eq.~(\ref{cp1}). In Fig.~\ref{couple}(a), we display the behavior of $R$ within the interval $\pi < \phi < 3\pi$. The figure demonstrates that all the curves exhibit symmetry about the frequency $\Omega d/v = 2\pi$. Away from this central point, the decay rate is significantly suppressed, especially for larger $M$. Consequently, the curves for $R$ take on a characteristic window shape with a width of $2\pi/M$.
More importantly, we identify the optimal working points where $|\partial R/\partial \Omega|$ reaches its maximum value. These points are marked by four stars in Fig.~\ref{couple}(a), and the corresponding frequency is denoted as $\Omega_{\rm opt}$ in the following discussion.

For setups with more coupling points, we plot $\Omega_{\rm opt}$ versus $M$ in Fig.~\ref{couple}(b) using empty circles. The plot shows that the optimal frequency approaches $\Omega_{\rm opt} \rightarrow 2\pi v/d$ as the number of coupling points $M$ tends to infinity. Numerical fitting reveals that $\phi_{\rm opt}/\pi = \Omega_{\rm opt} d/(v \pi) \approx 2 + M^{-1}$, which is depicted by the solid curve. This further confirms the window-shaped curves in Fig.~\ref{couple}(a), where the optimal working point lies approximately at the half-width of the window.
Additionally, the value of $|\partial R/\partial \Omega|$ is investigated, as shown by the orange dot dash line in Fig.~\ref{couple}(b). It follows a linear scaling with respect to the number of coupling points $M$, suggesting that increasing the number of coupling points will benefit the metrology.

\subsection{Beyond Markovian approximation}
\begin{figure}
  \centering
  \includegraphics[width=8cm]{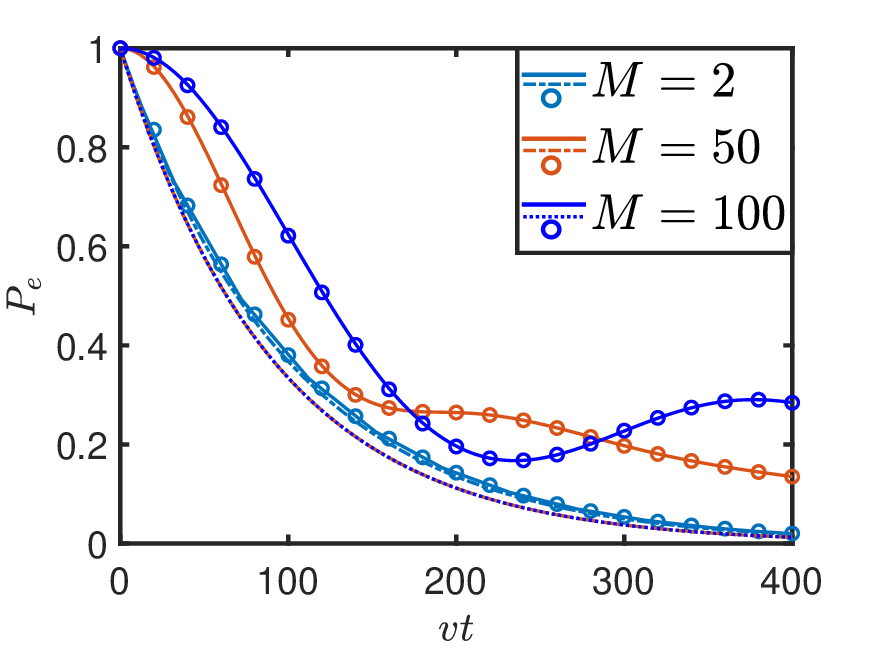}
  \caption{Dynamics of the giant emitter. Coupling strength $G=0.1V$. The dash line, empty circles and the solid line separately represent Markovian, numerical and analysis~(by Eq.~(\ref{dynamics1})) dynamics.  }\label{kapibala}
\end{figure}

In the above subsection, we identified the optimal working point for frequency metrology under the Markovian approximation. In Fig.~\ref{kapibala}, we further plot the population dynamics $P_e = \langle |e\rangle\langle e| \rangle$ of the initially excited giant emitter when coupled to the waveguide at $M$ coupling points at the corresponding optimal working point.
Under the Markovian approximation, the dot dash curves represent an exponential decay characterized by $P_e = \exp(-Rt)$. In addition, we plot the numerical results (empty circles) obtained from the unitary evolution governed by $|\psi(t)\rangle = \exp(-iHt)|\psi(0)\rangle$, where the initial state is $|\psi(0)\rangle = \sigma_+|g,\mathrm{vac}\rangle$. Here, $|g,\mathrm{vac}\rangle$ represents the emitter in its ground state while the waveguide is in the vacuum state.
For the numerical calculation, we impose a frequency cut-off for the waveguide around $\Omega_{\rm opt}$, restricting $|k|d \in (4\pi/3, 8\pi/3)$, and consider $200$ modes in the simulation.

Since we are working in the weak emitter-waveguide coupling regime, the numerical and analytical results agree perfectly for small $M$, such as $M = 2$. However, as $M$ increases, the two curves gradually deviate from each other, with the numerical results exhibiting an oscillatory behavior for larger $M$. This suggests that the retardation effect, arising from the nonlocal coupling between the emitter and the waveguide via multiple coupling points, breaks the Markovian approximation, as reported in previous studies~\cite{chengre,LG2017}.
To accurately capture the dynamics, it is necessary to go beyond the Markovian approximation. For this purpose, we assume the wave function of the system in the single-excitation subspace as

\begin{equation}
|\psi(t)\rangle=\left[\alpha(t)\sigma_++\int dk \beta_k(t)a_k^\dagger\right]|g,\rm{vac}\rangle
\end{equation}
where $\alpha$ and $\beta_k$ are respectively the excitation amplitudes of the emitter and $k$th mode in the waveguide.

As shown in Appendix~\ref{nonm}, the Sch\"{o}dinger equation $i\partial_t |\psi(t)\rangle=H|\psi(t)\rangle$ supplies us with the emitter's amplitude
\begin{eqnarray}
\dot{\alpha}(t)=&&-\frac{G^2}{Mv}\alpha(t)\nonumber\\
&&-\frac{2G^2}{M^2v}\sum_{l=1}^{M-1}(M-l)e^{i\Omega l\tau}\alpha(t-l\tau)\Theta(t-l\tau)\label{dynamics1}
\end{eqnarray}
where $\Theta(\bullet) $ is the Heaviside step function.
The first term represents the dissipation process at each individual coupling point. The second term accounts for the retardation effect, represented by the Heaviside step function, and the interference effect, represented by the factor $\exp(i\Omega l\tau)$ in the summation. These two effects result in a significant non-Markovian behavior, as $\alpha(t)$ depends on its historical values $\alpha(t-l\tau)$ for $l = 1, \dots, M-1$.
By neglecting these effects and approximating $\alpha(t-l\tau)\Theta(t-l\tau) \rightarrow \alpha(t)$, the dynamics reduce to the Markovian process described by $\dot{\alpha}(t) = -R\alpha(t)/2$, which is consistent with the result derived from the master equation. In Fig.~\ref{kapibala}, the result obtained from Eq.~(\ref{dynamics1}) is shown by the solid line, which agrees with the numerical results for both small and large values of $M$.

\section{Magnetometer}
\label{meter}
\begin{figure}[t]
  \centering
  \includegraphics[width=4.2cm]{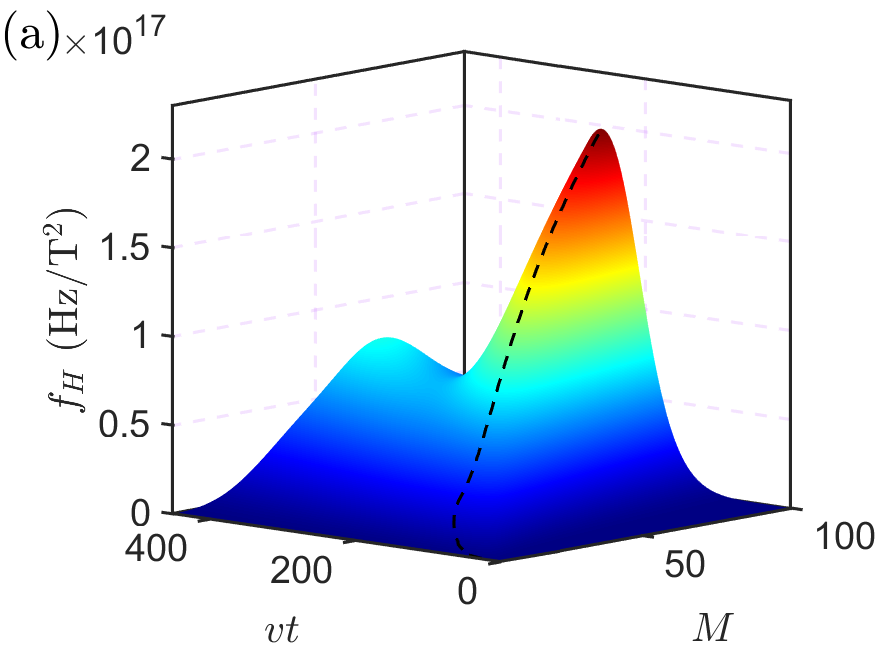} \includegraphics[width=4.2cm]{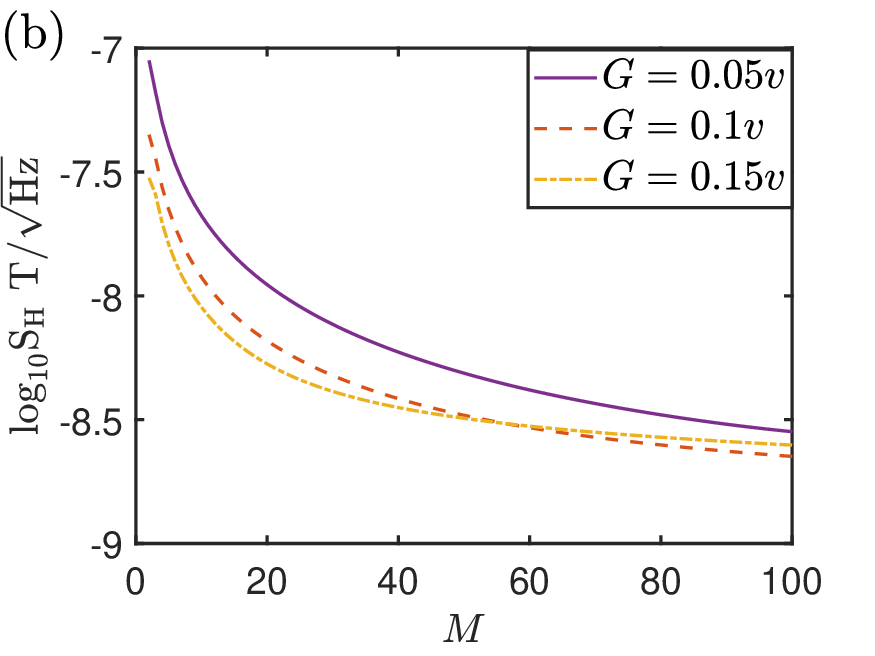}
  \caption{The variation of the CFI per unit time~(a) and sensitivity~(b) in the giant emitter setup. $G=0.1v$ for (a).}\label{sen1}
\end{figure}

In recent experimental advancements, the giant emitter has been realized using a spin ensemble in a YIG sphere~\cite{mq}, which couples to bent superconducting circuits acting as the waveguide with a linear dispersion relation. In principle, the Rabi oscillations between the two lowest states of the Kittel mode in the YIG sphere can be observed with the assistance of auxiliary quantum systems, such as a transmon qubit. This enables us to model the YIG sphere as a two-level emitter, as described in the previous section. Since the frequency of the Kittel mode is proportional to the bias MF $H$ via $\Omega = \gamma H$, with $\gamma = 175\,\mathrm{GHz/T}$ being the gyromagnetic ratio, this setup can be used to design a magnetometer.

The fluctuation of the MF is related to that of the frequency by \(\Delta^2 H=\Delta^2 \Omega/\gamma^2\), and CFI for the single time population measurement is $\mathcal{F}_H=1/(\Delta^2 H)$. Corresponding, the CFI per unit time for the proposed magnetometer reads
\begin{eqnarray}
f_{H}=\frac{\mathcal{F}_H}{t}=\frac{\gamma^2 t}{e^{Rt}-1}(\frac{\partial R}{\partial \Omega})^2\label{Fis}.
\end{eqnarray}
and the sensitivity is defined by $S_H=1/\sqrt{f_H}$, which has been widely applied in various quantum sensing and metrology schemes.

To analyze the CFI and sensitivity of our magnetometer proposal, we first choose the optimal working point demonstrated in Fig.~\ref{couple}(a) within the Markovian approximation. However, to ensure the validity of the metrology process, we track the system's dynamics using the numerical time evolution of \( |\psi(t)\rangle = \exp(-iHt)|g,\mathrm{vac}\rangle \). Although beyond the Markovian process, the population probability \( P_e \) can still be expressed in the form \( P_e = \exp(-Rt) \), where \( R \) becomes time-dependent.
As a result, the CFI per unit time \( f_H \) depends on both the evolution time and the number of coupling points \( M \), as shown in Fig.~\ref{sen1}(a). For a fixed \( M \), \( f_H \) exhibits a non-monotonic evolution over time. The maximum value of \( f_H \) for each \( M \), indicated by the dash dark curve, increases monotonically with \( M \). This suggests that a YIG sphere with more coupling points to the waveguide enhances the precision of the magnetometer. Notably, \( f_H \) can reach as high as \( 2 \times 10^{17}\,{\rm Hz/T^2} \).
Furthermore, the sensitivity \( S_H \) as a function of \( M \) is shown in Fig.~\ref{sen1}(b) for different coupling strengths \( g \). When \( M \) is sufficiently large, for instance \( M = 100 \), the sensitivity for magnetic field sensing achieves the order of \( 10^{-8} - 10^{-9}\,{\rm T/\sqrt{Hz}} \).

\section{Sensitivity in small emitters setup}
\label{small}

In the previous sections, we have proposed a magnetometer based on the giant emitter setup. To benchmark the advantage of the self-interference effect in enhancing the MF sensing, we compare the sensitivity with that of small emitters system.

\begin{figure}[t]
  \centering
  \includegraphics[width=6cm]{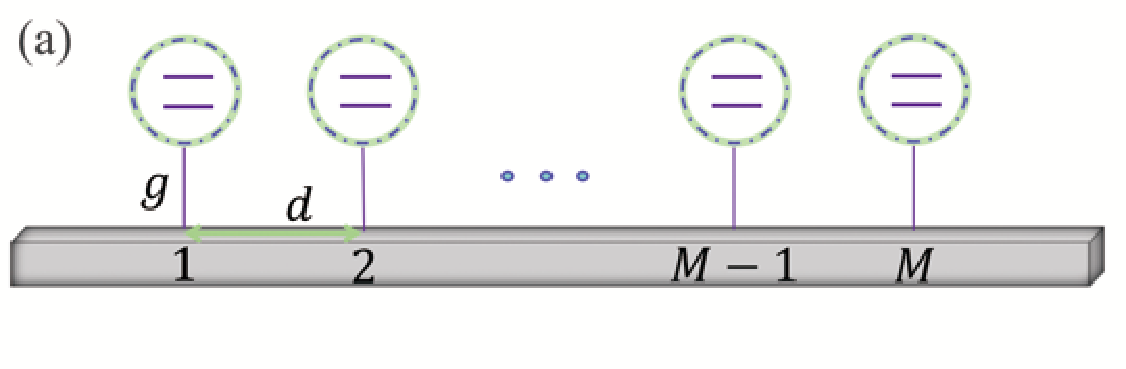}
  \includegraphics[width=4.2cm]{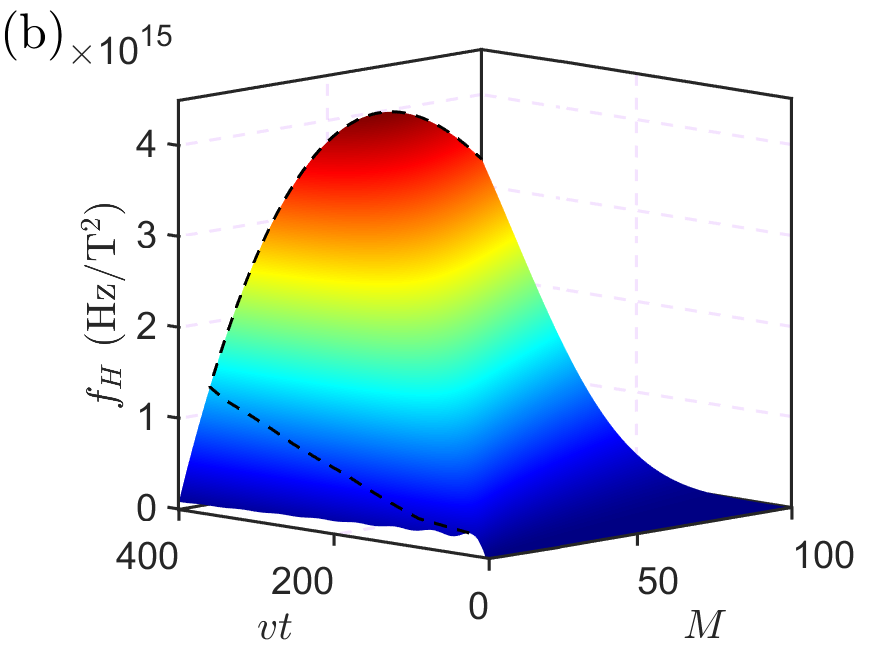} \includegraphics[width=4.2cm]{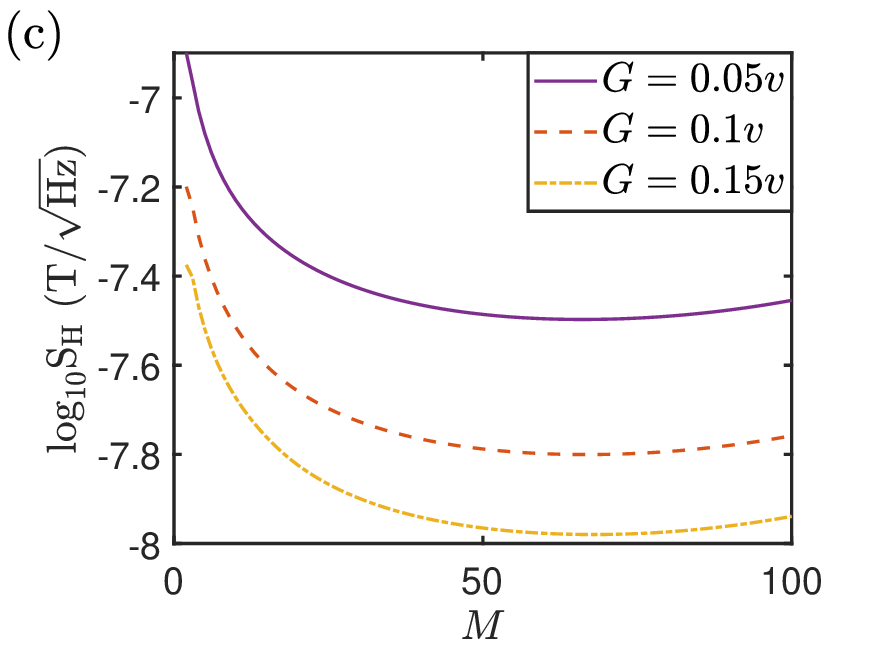}
  \caption{(a) $M$ small emitters are coupled to different positions of the waveguide. The distance between adjacent small atoms is $d$. (b,c) The variation of the CFI per unit time and sensitivity in small emitters setup. $G=0.1v$ for (b). }\label{sen2}
\end{figure}

As sketched in Fig.~\ref{sen2}(a), we replace the giant emitter in the previous setup by $M$ small emitters, with each one being located at the original coupling points. In such scheme, the photon in the waveguide can still be reflected by emitters, but the self-interference effect does not exist. The Hamiltonian of small emitters system is written as $H=H_s+H_W+H_{I,s}$ where
\begin{eqnarray}
H_s&=&\Omega \sum_{n=1}^{M}|e\rangle_n\langle e|,\nonumber\\
H_W&=&\int_{-\infty}^{+\infty} v |k|a_{k}^{\dagger}a_k dk,\nonumber\\
H_{I,s}&=&\frac{g}{\sqrt{2\pi}}\sum_{n=1}^{M}\int_{-\infty}^{+\infty}
dk\left(\sigma_+^{(n)}a_k e^{iknd}+{\rm H.c.}\right).
\end{eqnarray}

For the purpose of comparison, we assume that all small emitters have the same transition frequency \( \Omega \) and coupling strength \( g \) to the waveguide, which are equal to those in the previous giant emitter setup. Additionally, we restrict the system to the single-excitation subspace by preparing the initial state as $|\psi(0)\rangle=\sum_{i=1}^{M} \sigma_+^{(i)}|g,{\rm vac}\rangle/\sqrt{M}$,
where \( M \) emitters are in a single-excitation Dicke (entangled) state, and the waveguide is in the vacuum state. We choose the physical quantity $P_e^{(s)}(t)=\sum_n\langle |e\rangle_{n}\langle e|\rangle$ as the observable to discuss the CFI per unit time and sensitivity, which are plotted in Fig.~\ref{sen2}(b) and (c), respectively.
In comparison to the giant emitter setup, we find that the CFI per unit time is $100$ times lower, while the optimal sensitivity only reaches $10^{-7}-10^{-8}\,{\rm T/\sqrt{Hz}}$. These results imply that the self-interference in the giant emitter based magnetometer is superior than the small ones even the entanglement as the quantum source has to be prepared in the later setup.

\section{Conclusions}
\label{con}

In summary, we find that under the Markovian approximation, the relaxation rate of the giant emitter exhibits a linear dependence on the number of coupling points in the multi-point coupled configuration of the YIG sphere and the bent waveguide. Leveraging this property, we derive the measurement sensitivity for the magnetic field strength, which controls the frequency of the emitter, by observing the population probability of the excited state of the giant emitter.

We have thoroughly analyzed the dependence of the CFI per unit time on both the evolution time and the number of coupling points \(M\). Our results reveal that the maximum CFI per unit time scales linearly with \(M\), and the measurement sensitivity can reach \(10^{-8} - 10^{-9}\,{\rm T/\sqrt{Hz}}\) when \(M\) is large.

These findings are attributed to the self-interference effect of the giant emitter. To benchmark this advantage, we applied the same measurement scheme to a system composed of multiple small emitters. The results show that the measurement sensitivity of the small emitter system is an order of magnitude lower than that of the giant emitter system. This highlights the superiority of the self-interference effect in the giant emitter configuration over the interference effects in small emitter systems for precision measurements.

\section*{Acknowledgments}

This work is supported by the funding from Jilin Province (Grant Nos. 20230101357JC), the National Science Foundation of China (Grant No. 12375010) and the Innovation Program for Quantum Science and Technology (No. 2023ZD0300700).

\appendix%\appendixpage
\addcontentsline{toc}{section}{Appendices}\markboth{APPENDICES}{}
\begin{subappendices}
\begin{widetext}
\section{Markov master equation}\label{mas}
Considering the waveguide as the environment, the dynamics of the giant emitter can be described by a master equation. In the interaction picture, the interaction Hamiltonian is given by

\begin{eqnarray}
V_I(t)=\frac{g}{\sqrt{2\pi}}\sum_{n=1}^{M}\int_{-\infty}^{\infty}dk\left(\sigma_+a_k e^{i(\Omega-v|k|)}e^{iknd}+\rm{H.c.}\right).
\end{eqnarray}

Under the Markov approximation, the formal master equation for a quantum open system reads

\begin{eqnarray}
\dot\rho(t)=-\int_{0}^{\infty}d\tau {\rm Tr}_c\left[V_I(t),\left[V_I(t-\tau),\rho_c\otimes\rho(t)\right] \right],
\end{eqnarray}

Initially, the waveguide is in the vacuum state at the zero temperature. The above equation can be reduced as (go back to the Schr\"{o}dinger picture)
\begin{equation}
\dot{\rho}=-i\Omega[|e\rangle\langle e|,\rho]+(A+A^{*})\sigma^{-}\rho\sigma^{+}
-A\sigma^{+}\sigma^{-}\rho-A^{*}\rho\sigma^{+}\sigma^{-},
\end{equation}
where

\begin{eqnarray}
A&=&\frac{g^2}{2\pi}\sum_{n,n'=1}^{M}\int_{0}^{\infty}d\tau\int_{-\infty}^{\infty}dke^{i(\Omega-v|k|)\tau}e^{ik(n-n')d}\nonumber\\
&=&\frac{g^2}{2\pi}\sum_{n,n'=1}^{M}\int_{0}^{\infty}d\tau e^{i\Omega \tau}\left[\int_{0}^{\infty}dk e^{-ikv(\tau-\frac{(n-n')d}{v})}+\int_{0}^{\infty}dk e^{-ikv(\tau+\frac{(n-n')d}{v})}\right]\nonumber\\
&=&\frac{g^2}{2\pi}\sum_{n,n'=1}^{M}\int_{0}^{\infty}d\tau e^{i\Omega \tau}\frac{1}{v}\left[\int_{0}^{\infty}d\omega_k e^{-i\omega_k[\tau-\frac{(n-n')d}{v}]}+\int_{0}^{\infty}d\omega_k e^{-i\omega_k(\tau+\frac{(n-n')d}{v})}\right]\nonumber\\
&=&\frac{g^2}{2\pi}\sum_{n,n'=1}^{M}\int_{0}^{\infty}d\tau e^{i\Omega \tau}\frac{1}{v}\left[\int_{-\infty}^{\infty}d\omega_k e^{-i\omega_k[\tau-\frac{(n-n')d}{v}]}+\int_{-\infty}^{\infty}d\omega_k e^{-i\omega_k(\tau+\frac{(n-n')d}{v})}\right]\nonumber\\
&=&\frac{g^2}{v}\sum_{n,n'=1}^{M}\int_{0}^{\infty}d\tau e^{i\Omega \tau}\left(\delta[\tau-\frac{(n-n')d}{v}]+\delta[\tau+\frac{(n-n')d}{v}]\right)\nonumber\\
&=&\frac{g^2}{v}\left(\sum_{n>n'}e^{\frac{i\Omega(n-n')d}{v}}+\sum_{n<n'}e^{\frac{i\Omega(n'-n)d}{v}}\right)\nonumber\\
&=&\frac{g^2}{v}\sum_{n,n'=1}^{M}e^{\frac{i\Omega|n-n'|d}{v}}
\end{eqnarray}
Here, $\omega_k=v|k|$ and we use analytic extension to derive the above equation. Now, we can get the decay rate and Lamb shift
\begin{eqnarray}
R&=&2\text{Re}(A)=\frac{2g^2}{v}\sum_{n,l=1}^{M}\cos [|n-l|\phi]\\
L&=&\text{Im}(A)=\frac{g^2}{v}\sum_{n,l=1}^M\sin [|n-l|\phi]
\end{eqnarray}
where $\phi=\Omega d/v$ is accumulated by photons between adjacent coupling points of the giant emitter.

\section{Non-Markov dynamics}\label{nonm}
The wave function of the system in the single excitation subspace can be assumed as
\begin{equation}
|\psi(t)\rangle=e^{-i\Omega t}\left[\alpha(t)\sigma_++\int dk \beta_k(t)a_k^\dagger\right]|g,\rm{vac}\rangle
\end{equation}

According to the Sch\"{o}dinger equation $i\partial_t |\psi(t)\rangle=H|\psi(t)\rangle$ , we can get
\begin{eqnarray}
i\dot\alpha(t)&=&\frac{g}{\sqrt{2\pi}}\sum_{n=1}^{M}\int_{-\infty}^{\infty}dk \beta_k(t)e^{iknd},\label{alpha}\\
i\dot{\beta_k}(t)&=&(\omega_k-\Omega)\beta_k+\frac{g}{\sqrt{2\pi}}\sum_{n=1}^{M}\alpha(t)e^{-iknd}.\label{beta}
\end{eqnarray}
where $\omega_k=v|k|$ is the frequency of the waveguide. Then, under the initial condition $\alpha(0)=1$ and $\beta_k(0)=0$, Eq.~(\ref{beta}) can be integrated as
\begin{eqnarray}
\beta_k(t)=-i\frac{g}{\sqrt{2\pi}}\sum_{n=1}^{M}\int_{0}^{t}d\tau\alpha(\tau)e^{-iknd}e^{-i(\omega_k-\Omega)(t-\tau)}.
\end{eqnarray}

Take the above equation into the Eq.~(\ref{alpha}),
\begin{eqnarray}
\dot\alpha(t)&=&-\frac{g^2}{{2\pi}}\sum_{n,n'=1}^{M}\int_{-\infty}^{\infty}dk \int_{0}^{t}dt_1\alpha(t_1)e^{ik(n-n')d}e^{-i(\omega_k-\Omega)(t-t_1)}\nonumber\\
&=&-\frac{g^2}{{2\pi}v}\sum_{n,n'=1}^{M} \int_{0}^{t}dt_1 \alpha(t_1)e^{i\Omega(t-t_1)}\left[\int_{0}^{\infty}d\omega_k \left(e^{i\omega_k\frac{(n-n')d}{v}}+e^{-i\omega_k\frac{(n-n')d}{v}}\right)e^{-i\omega_k(t-t_1)}\right]\nonumber\\
&=&-\frac{g^2}{{2\pi}v}\sum_{n,n'=1}^{M} \int_{0}^{t}dt_1 \alpha(t_1)e^{i\Omega(t-t_1)}\left[\int_{-\infty}^{\infty}d\omega_k \left(e^{i\omega_k\frac{(n-n')d}{v}}+e^{-i\omega_k\frac{(n-n')d}{v}}\right)e^{-i\omega_k(t-t_1)}\right]\nonumber\\
&=&-\frac{g^2}{v} \int_{0}^{t}dt_1 \alpha(t_1)e^{i\Omega(t-t_1)}\left(\sum_{n>n'}\delta\left[\tau_{n,n'}-(t-t_1)\right]+\sum_{n<n'}\delta\left[\tau_{n',n}-(t-t_1)\right]\right)\nonumber\\
&=&-\frac{g^2}{v} \int_{0}^{t}dt_1 \alpha(t-t_1)e^{i\Omega t_1}\left(\sum_{n>n'}\delta\left[\tau_{n,n'}-t_1\right]+\sum_{n<n'}\delta\left[\tau_{n',n}-t_1\right]\right)\nonumber\\
&=&-\frac{g^2}{v}\sum_{n,n'=1}^{M}\alpha(t-|\tau_{n,n'}|)e^{i\Omega|\tau_{n,n'}|}\Theta\left(t-|\tau_{n,n'}|\right)\nonumber\\
&=&-\frac{Mg^2}{v}\alpha(t)-\frac{2g^2}{v}\sum_{l=1}^{M-1}(M-l)\alpha(t-l\tau)e^{i\Omega l\tau}\Theta\left(t-l\tau\right)
\end{eqnarray}
where $\tau_{n,m}=(n-m)d/v$ and $\tau=d/v$.

\end{widetext}
\end{subappendices}


\begin{thebibliography}{99}

\bibitem{ge}Rachael Parker, Alastair Ruffell, David Hughes, and Jamie Pringle, Geophysics and the search of freshwater bodies: A review, Sci. Justice {\bf50}, 141 (2010).

\bibitem{ge1}H. B. Dang, A. C. Maloof, and M. V. Romalis, Ultrahigh sensitivity magnetic field and magnetization measurements with an atomic magnetometer, Appl. Phys. Lett. {\bf97}, 151110 (2010).

\bibitem{tian}J. S. Bennett, B. E. Vyhnalek, H. Greenall, E. M. Bridge, F. Gotardo, S. Forstner, G. I. Harris, F. A. Miranda, and W. P. Bowen, Precision magnetometers for aerospace applications: A review, Sensors {\bf21}, 5568 (2021).

\bibitem{biomedical}D. Murzin, D. J. Mapps, K. Levada, V. Belyaev, A. Omelyanchik, L. Panina, and V. Rodionova, Ultrasensitive magnetic field sensors for biomedical applications,
Sensors {\bf20}, 1569 (2020).

\bibitem{OMr}D. Budker, and M. Romalis, Optical magnetometry, Nat. Phys. {\bf 3}, 227 (2007).

\bibitem{OMr3}F. Wolfgramm, A. Cer\`{e}, F. A. Beduini, A. Predojevi\'{c}, M. Koschorreck, and M. W. Mitchell, Squeezed-Light Optical Magnetometry,
Phys. Rev. Lett.  {\bf 105}, 053601 (2010).

\bibitem{OMr2}B. Patton, E. Zhivun, D. C. Hovde, and D. Budker, All-Optical Vector Atomic Magnetometer, Phys. Rev. Lett.  {\bf 113}, 013001 (2015).

\bibitem{OMr1}A. M. Akulshin, D. Budker, F. P. Bustos, T. Dang, E. Klinger, S. M. Rochester, A. Wickenbrock, and R. Zhang, Remote detection optical magnetometry, Phys. Rep. {\bf 1106}, 1 (2025).

\bibitem{S1}M. V. Gustafsson, T. Aref, A. F. Kockum, M. K. Ekstr\"om, G. Johansson, and P. Delsing, Propagating phonons coupled to an artificial atom, Science {\bf346}, 207 (2014).

\bibitem{AF2018}A. F. Kockum, G. Johansson and F. Nori, Decoherence-free interaction between giant atoms in waveguide quantum electrodynamics, Phys. Rev. Lett.  {\bf 120}, 140404 (2018).

\bibitem{AC2020}A. Carollo, D. Cilluffo and F. Ciccarello, Mechanism of decoherence-free coupling between giant atoms, Phys. Rev. Research {\bf 2}, 043184 (2020).

\bibitem{DL2023PRA}L. Du, L. Guo, and Y. Li, Complex decoherence-free interactions between giant atoms, Phys. Rev. A \textbf{107}, 023705 (2023).

\bibitem{Lambpra14}A. F. Kockum, P. Delsing, and G. Johansson, Designing frequency-dependent relaxation rates and Lamb shifts for a giant artificial atom, Phys. Rev. A \textbf{90}, 013837 (2014).

\bibitem{ar2022}S. Terradas-Brians\'o, C. A. Gonz\'alez-Guti\'errez, F. Nori, L. Mart\'{\i}n-Moreno, and D. Zueco, Ultrastrong waveguide QED with giant atoms, Phys. Rev. A {\bf 106}, 063717 (2022).

\bibitem{NM1}L. Guo, A. F. Kockum, F. Marquardt, and G. Johansson, Oscillating bound states for a giant atom, Phys. Rev. Research \textbf{2}, 043014 (2020).

\bibitem{NM2}C. A. Gonz\'{a}lez-Guti\'{e}rrez, J. Rom\'{a}n-Roche, and D. Zueco, Distant emitters in ultrastrong waveguide QED: Ground-state properties and non-Markovian dynamics, Phys. Rev. A \textbf{104}, 053701 (2021).

\bibitem{kim2023}K. H. Lim, W.-K. Mok, and L.-C. Kwek, Oscillating bound states in non-Markovian photonic lattices, Phys. Rev. A \textbf{107}, 023716 (2023).

\bibitem{guonjp}L. Xu, and L. Guo, Catch and release of propagating bosonic field with non-Markovian giant atom, New J. Phys. \textbf{26}, 013025 (2024).

\bibitem{XW2021}X. Wang, T. Liu, A. F. Kockum, H.-R. Li, and F. Nori, Tunable chiral bound states with giant atoms,
 Phys. Rev. Lett. {\bf 126}, 043602 (2021).

\bibitem{XW2022q}X. Wang, and H.-R. Li, Chiral quantum network with giant atoms, Quantum Sci. Technol. {\bf7} 035007 (2022).

\bibitem{AS2022}A. Soro, and A. F. Kockum, Chiral quantum optics with giant atoms, Phys. Rev. A {\bf 105}, 023712 (2022).

\bibitem{DL2023}L. Du, Y.-T. Chen, Y. Zhang, Y. Li, and J.-H. Wu, Decay dynamics of a giant atom in a structured bath with broken
time-reversal symmetry, Quantum Sci. Technol. {\bf8}, 045010 (2023).

\bibitem{prx2023}C. Joshi, F. Yang, and M. Mirhosseini, Resonance Fluorescence of a Chiral Artificial Atom, Phys. Rev. X {\bf 13}, 021039 (2023).

\bibitem{FR2024}F. Roccati and D. Cilluffo, Controlling Markovianity with Chiral Giant Atoms, Phys. Rev. Lett. {\bf 133}, 063603 (2024).

\bibitem{guo}X. Guo, and Z. Wang, Controllable superradiance scaling in photonic waveguide, arXiv: 2501.02806 (2025).

\bibitem{Jieqiao} X.-L. Yin, W.-B. Luo, and J.-Q. Liao, Non-Markovian disentanglement dynamics in double-giant-atom waveguide-QED systems, Phys. Rev. A \textbf{106}, 063703 (2022).

\bibitem{Jieqiao1}X.-L. Yin and J.-Q. Liao, Generation of two-giant-atom entanglement in waveguide-QED systems, Phys. Rev. A \textbf{108}, 023728 (2023).

\bibitem{PRL2023}A. C. Santos and R. Bachelard, Generation of Maximally Entangled Long-Lived States with Giant Atoms in a Waveguide, Phys. Rev. Lett. {\bf 130}, 053601 (2023).

\bibitem{weng}M. Weng, Xin Wang, and Z. Wang, Interaction and entanglement engineering in a driven-giant-atom setup with a coupled resonator waveguide, Phys. Rev. A \textbf{110}, 023721 (2024).

\bibitem{LG2017}L. Guo, A. Grimsmo, A. F. Kockum, M. Pletyukhov, and G.
Johansson, Giant acoustic atom: A single quantum system with a deterministic time delay, Phys. Rev. A {\bf 95}, 053821 (2017).

\bibitem{chengre}W. Cheng, Z. Wang, and Y.-X. Liu, Topology and retardation effect of a giant atom in a topological waveguide, Phys. Rev. A {\bf 106}, 033522 (2022).

\bibitem{sq1}G. Andersson, B. Suri, L. Guo, T. Aref, and P. Delsing, Non-exponential decay of a giant artificial atom, Nat. Phys. \textbf{15}, 1123 (2019).

\bibitem{sq2}B. Kannan, M. J. Ruckriegel, D. L. Campbell, A. F. Kockum, J. Braum\"uller, D. K. Kim, M. Kjaergaard, P. Krantz, A. Melville, B. M. Niedzielski, A. Veps\"al\"ainen, R. Winik, J. L. Yoder, F. Nori, T. P. Orlando, S. Gustavsson, and W. D. Oliver, Waveguide quantum electrodynamics with superconducting artificial giant atoms, Nature \textbf{583}, 775 (2020).

\bibitem{sq3}A. M. Vadiraj, A. Ask, T. G. McConkey, I. Nsanzineza, C. W. Sandbo Chang, A. F. Kockum, and C. M. Wilson, Engineering the level structure of a giant artificial atom in waveguide quantum electrodynamics, Phys. Rev. A, {\bf 103}, 023710 (2021).

\bibitem{sq4}J. Hu, D. Li, Y. Qie, Z. Yin, A. F. Kockum, F. Nori, and S. An, Engineering the Environment of a Superconducting Qubit with an Artificial Giant Atom, arXiv: 2410, 15377 (2024).

\bibitem{mq}Z.-Q. Wang, Y.-P. Wang, J. Yao, R.-C. Shen, W.-J. Wu, J. Qian, J. Li, S.-Y. Zhu, and J. Q. You, Giant spin ensembles in waveguide magnonics, Nat. Commun. {\bf 13}, 7580 (2022).

\bibitem{smq1}Y. Tabuchi, S. Ishino, A. Noguchi, T. Ishikawa, R. Yamazaki, K. Usami, and Y. Nakamura, Coherent coupling between a ferromagnetic magnon and a superconducting qubit, Science {\bf349}, 6246 (2015).

\bibitem{smq2}D. Lachance-Quirion, S. P. Wolski, Y. Tabuchi, S. Kono, K. Usami, and Y. Nakamura, Entanglement-based single-shot detection of a single magnon with a superconducting qubit, Science {\bf367}, 6476 (2020).

\bibitem{smq4}S. P. Wolski, D. Lachance-Quirion, Y. Tabuchi, S. Kono, A. Noguchi, K. Usami, and Y. Nakamura, Dissipation-Based Quantum Sensing of Magnons with a Superconducting Qubit, Phys. Rev. Lett. {\bf125}, 117701 (2020).

\bibitem{smq3}D. Xu, X.-K. Gu, H.-K. Li, Y.-C. Weng, Y.-P. Wang, J. Li, H. Wang, S.-Y. Zhu, and J. Q. You, Quantum Control of a Single Magnon in a Macroscopic Spin System, Phys. Rev. Lett. {\bf130}, 193603 (2023).

\bibitem{CG2024}C.-G. Liu, C.-W. Lu, N.-N. Zhang, and Q. Ai, Quantum simulation of bound-state-enhanced quantum metrology, Phys. Rev. A {\bf 109}, 042623 (2024).

\end{thebibliography}
\end{document}